# Information-technology approach to quantum feedback control

## Dao-Yi Dong*, Chen-Bin Zhang and Zong-Hai Chen

Department of Automation, University of Science and Technology of China, Hefei 230027,
People's Republic of China
∗E-mail: dydong@mail.ustc.edu.cn

**Abstract**
Quantum control theory is profitably reexamined from the perspective of quantum information, two results on the role of quantum information technology in quantum feedback control are presented and two quantum feedback control schemes, teleportation-based distant quantum feedback control and quantum feedback control with quantum cloning, are proposed. In the first feedback scheme, the output from the quantum system to be controlled is fed back into the distant actuator via teleportation to alter the dynamics of system. The result theoretically shows that it can accomplish some tasks such as distant feedback quantum control that Markovian or Bayesian quantum feedback can't complete. In the second feedback strategy, the design of quantum feedback control algorithms is separated into a state recognition step, which gives "on-off" signal to the actuator through recognizing some copies from the cloning machine, and a feedback (control) step using another copies of cloning machine. A compromise between information acquisition and measurement disturbance is established, and this strategy can perform some quantum control tasks with coherent feedback.

## 1. Introduction

Controlling quantum phenomena is a most important problem that lies at the heart of several fields including atomic physics, molecular chemistry and quantum information [1]. Especially in the rapidly developing quantum information theory, since information is represented with quantum bit and quantum information processing is essentially the manipulation of quantum state, controlling quantum system to desired state is an important and key task. So quantum control theory rapidly develops in recent years. Current research of quantum control mainly involves controllability of quantum system [2-8], coherent control [9-11], quantum optimal control [12] and quantum feedback control [13-16]. Some methods of geometric control and coherent control have been applied in the fields of nuclear magnetic resonance and selective-bond chemistry. With the development of quantum control theory, it is also applied to quantum computation and quantum communication to protect quantum information from noise and decoherence.

On the one hand, the emergence of quantum control theory partly comes from the need of quantum information. On the other hand, the rapid development and gradual maturation of quantum information technology suggest that quantum control theory might profitably be reexamined from the perspective of quantum information [17]. Therefor, Lloyd has proposed coherent quantum feedback scheme [14], in which the sensors, controllers, and actuators are quantum systems that interact coherently with the system to be controlled. Since the controllers can process and feed back quantum information, Lloyd's feedback method can perform some tasks such as entanglement transfer that are not possible using quantum control with classical feedback. Recently, Kawabata has shown the information-theoretic analysis of quantum open-loop and feedback control, and has also given the fundamental limits on the control of quantum mechanical systems from the viewpoint of information theory [18]. By applying the theoretic results to an entropic analysis of quantum error correcting code, Kawabata has shown that the quantum error correction procedure can be regarded as an information theoretically optimal quantum feedback control system.

With the proliferation of results on quantum information, it is possible to use more information technologies to quantum control theory. This paper presents two results on the role of quantum information technology in quantum feedback control and proposes two quantum feedback control schemes: one is teleportation-based distant quantum feedback control and the other is quantum feedback control with quantum cloning. The following section gives a brief introduction to quantum control theory and quantum feedback control. Sec.3 and Sec.4 present two quantum feedback control strategies using quantum information technology, respectively. Conclusion is given in Sec.5.

## 2. Quantum control and quantum feedback control

### 2.1 Quantum control theory

Quantum control theory is becoming a rapidly increasing research field and it expects to determine how to drive quantum mechanical systems from an initial given state to a pre-determined target state with some given time T [19]. In quantum control, the systems to be controlled are mainly some specific quantum systems such as atoms, spin-1/2 electrons, Bose-Einstein condensates, polarization photons, etc., whose behavior does not admit an accurate classical description. The current main fields of quantum control are controllability, quantum optimal control, coherent control and closed-loop control.

The controllability answers whether a controller can drive a quantum system to a desired state. The main tools are Lie group, Lie algebra and graph theory. Some necessary and sufficient conditions for some finite-dimensional Hamiltonian quantum systems have been obtained. Quantum optimal control theory is applied to find the best control fields to achieve the desired target. In optimal problem, selecting of cost function and searching of optimal control are two key steps. Under some specific circumstances, one can select the maximum (or minimum) of field energy, control time, expectation value, or their combination. Only to some simple examples, the exact analytic solutions of optimal control can be obtained. To general quantum optimal problem, we may apply numerical method and associated algorithms to search approximate optimal control fields. In quantum control, coherent control is a particularly powerful method. It coherently manipulates the state of a quantum system to the desired target state by applying semiclassical potentials. The result of coherent control is derived from the constructive and destructive quantum mechanical interference. Coherent control has been widely applied to chemical reaction, especially selectively breaking and making chemical bonds in polyatomic molecules.

The application of coherent control initially only concerned on quantum open-loop control [10]. With the development of quantum control theory, quantum closed-loop control has gradually been attached importance to. There are two general forms of quantum closed-loop controls [1]: closed learning control and quantum feedback control. Quantum feedback control follows the same sample from initiation of control to the final target; however, closed-loop learning control involves a closed-loop operation where each cycle of the loop is executed with a new sample. The procedure of learning control generally includes three elements: (i) an input trial control design; (ii) the laboratory generation of the control that is applied to the sample and subsequently observed for its impact, and (iii) a learning algorithm that considers the prior experiments and suggests the form of the next control for an excursion around the loop once again, and so on. Learning control has many successful applications in chemical reaction. At the same time, quantum feedback control develops in parallel with learning control and it has been widely applied in quantum optics and atomic physics.

**2.2 Quantum feedback control**

In classical control applications, feedback is a most effective strategy and recently it is also used to quantum control. Scientists expect to obtain information about the system from the quantum system to be controlled, process the information and feed it back to the system to complete active control of quantum system in a desired way. To control a system, one must obtain the information about the evolving system state through measurement. However, a quantum system is essentially different from a classical system. The measurement on a classical system doesn't change its state, but commonly a measurement of a quantum system will disturb its quantum state. Scientists have proposed some quantum feedback strategies including Markovian quantum feedback [16], Bayesian quantum feedback [20], non-Markovian quantum feedback with time delay [21] and coherent quantum feedback [14]. The first theoretical work on quantum feedback was presented by Yamamoto *et al.*[22], where they treated the fluctuations of the photocurrent in negative feedback way to generate amplitude squeezed state.

In 1993, Wiseman and Milburn first presented a quantum theory of optical feedback via homodyne detection, where the homodyne photocurrent was fed back onto optical cavity to alter the dynamics of the source cavity. In this quantum feedback theory, only the current photocurrent is immediately fed back and may then be forgotten, so the master equation describing the resulting evolution is Markovian and the theory is called Markovian quantum feedback [23]. It has been applied to generate squeezed light [16] and stabilize the internal state of atoms [24].

Markovian quantum feedback doesn't use the previous knowledge about the system to be controlled, therefor Doherty and Jacobs presented a quantum feedback scheme using continuous state estimation in 1999. They made the best of the detailed information from measurement record, and divided quantum feedback control process into two steps: a state estimation step and a feedback control step. Since the best state estimation will use all previous measurement results, not just the latest ones, the feedback is called Bayesian quantum feedback [23]. Recently, Doherty *et al.* also tried to consider separately optimizing state estimation step and feedback control step [13]. Bayesian quantum feedback has been used to cool and confine a single quantum degree of freedom [20], switch the state of a particle in a double-well potential [15] and control the decoherence of solid-state qubit [25]. Comparison research has shown that Bayesian quantum feedback is never inferior, and is usually superior, to Markovian quantum feedback in stabilizing the quantum state of the simplest nonlinear quantum system [23].

Both Markovian quantum feedback and Bayesian quantum feedback ignore the effect of feedback time delay. However, the delay can't be ignored in some situations. When nonzero feedback time delay is considered, the dynamics of system evolution exhibits strong non-Markovian. Giovannetti and co-workers first studied the non-Markovian quantum feedback with time delay in 1999. Their results show that feedback can also improve the dynamics of quantum systems for the delay time not too large.

Markovian quantum feedback, Bayesian quantum feedback and non-Markovian quantum feedback with time delay all use the feedback information from measurement results, however, measurement destroys the quantum characteristics of feedback information, so feedback information becomes classical information. As a result, although the system under control is quantum system, feedback controller processes classical information and effective quantum channel in feedback loop isn't constructed, so the strategies can be called quantum control with classical feedback. Differently, Lloyd presented a coherent quantum feedback scheme in which controller obtained quantum information, processed it and coherently fed back to the system to be controlled [14,26]. In this feedback strategy, the quantum characteristics in feedback loop aren't destroyed and the strategy can accomplish some tasks which are not possible using classical feedback. Moreover, Ting also proposed an alternative method for quantum feedback control, where a cloning machine served to obtain the feedback signal and the output [27]. Ting's method can also perform coherent feedback control at the cost of feeding precisely back the output. In this paper, we reexamine quantum feedback control from the perspective of quantum information technology and use quantum teleportation and quantum cloning in quantum information theory to design feedback channel.

## 3. Teleportation-based distant quantum feedback

In this section, we propose an alternative quantum feedback scheme, in which quantum teleportation is used to construct feedback controllers and the output from the quantum system to be controlled is fed back into the actuator via teleportation to alter the dynamics of system. Besides the tasks that Lloyd's quantum feedback can perform, this scheme can also accomplish the distant quantum feedback control task.

In the present scheme, quantum teleportation plays a key role in obtaining and transmitting feedback information. Teleportation, first proposed by Bennett *et al.*[28], is a process, in which an unknown quantum state can be disembodiedly transported to a desired receiver through purely classical communication and purely nonclassical Einstein-Podolsky-Rosen (EPR) correlations. A sender (Alice) is given an unknown state and also has one of two particles prepared in an EPR state. She performs a Bell-state measurement on the composite system of the unknown state and her EPR particle, and informs the receiver (Bob) the result of this measurement through classical communication. Knowing this, Bob performs one of four possible unitary transformations on the second EPR particle and converts it into an exact replica of the unknown state [28]. Experimental demonstration of quantum teleportation was realized with parametric down-conversion [29-31] and squeezed-state entanglement [32] in optical system. During the process, Alice will destroy the quantum state at hand and disembodiedly transmit it to Bob through two channels. So quantum teleportation can be viewed as a feedforward process. Here we will use teleportation as a feedback process and look it upon as a part of quantum feedback control loop.

The general picture of teleportation-based quantum feedback control is as follows. The quantum system to be controlled is called control object $S$. The actuator generates input signal to drive $S$ and the state $|\psi\rangle_S$ of the control object $S$ is given to Alice who is near to the output of system. In order to transmit the state $|\psi\rangle_S$ to Bob who is at the side of the actuator, a pair of particles $A$ and $B$ are prepared in an EPR state beforehand that Alice has particle $A$ and Bob has particle $B$. After obtaining the state

$|\psi\rangle_S$, Alice performs a Bell-state measurement on the combined system of the state $|\psi\rangle_S$ and particle *A*, and sends the result of this measurement via a classical channel to Bob. Depending on this result, Bob performs one of four possible unitary transformations. Now the state $|\psi\rangle_B$ of particle *B* lies in $|\psi\rangle_S$ and Alice's state $|\psi\rangle_S$ at hand is destroyed. After some appropriate transformations, the state $|\psi\rangle_B$ coherently interacts with the actuator, and the actuator generates new quantum signal to drive the control object again and starts next cycle until the control object *S* attains the desired target. In the process, quantum teleportation performs the distant transmission, so we can look upon it as a part of quantum feedback controller.

Without loss of generality, suppose the output state of control object is a simple coherent superposition state, i.e.

$$|\psi\rangle_S = \alpha|\uparrow\rangle_S + \beta|\downarrow\rangle_S \tag{1}$$

where $\alpha$ and $\beta$ are two complex numbers satisfying $|\alpha|^2 + |\beta|^2 = 1$, $|\uparrow\rangle_S$ and $|\downarrow\rangle_S$ represent spin up and spin down, respectively. The objective of quantum control is to put it into the state $|\uparrow\rangle$. In traditional control, the controller first makes a quantum measurement on the spin state, then if the result of the measurement is $|\uparrow\rangle$, do nothing, while if the result of the measurement is $|\downarrow\rangle$, put the spin in a static magnetic field and apply an electromagnetic pulse to flip the quantum spin state into the state $|\uparrow\rangle$ as desired. According to quantum measurement theory, this measurement procedure destroys the state of control object to be controlled. In the teleportation-based quantum feedback, firstly we prepare an entangled particle pair *A* and *B*, and the particle *A* and particle *B* are in an EPR state:

$$|\Psi^-\rangle_{AB} = \frac{1}{\sqrt{2}}(|\uparrow\rangle_A|\downarrow\rangle_B - |\downarrow\rangle_A|\uparrow\rangle_B) \tag{2}$$

So the complete state of *A*, *B* and *S* before Alice's measurement is:

$$|\psi\rangle_{SAB} = \frac{\alpha}{\sqrt{2}}(|\uparrow\rangle_S|\uparrow\rangle_A|\downarrow\rangle_B - |\uparrow\rangle_S|\downarrow\rangle_A|\uparrow\rangle_B) + \frac{\beta}{\sqrt{2}}(|\downarrow\rangle_S|\uparrow\rangle_A|\downarrow\rangle_B - |\downarrow\rangle_S|\downarrow\rangle_A|\uparrow\rangle_B) \tag{3}$$

Then Alice performs a measurement in the Bell basis vectors:

$$|\Psi^\pm\rangle_{SA} = \frac{1}{\sqrt{2}}(|\uparrow\rangle_S|\downarrow\rangle_A \pm |\downarrow\rangle_S|\uparrow\rangle_A) \tag{4}$$

$$|\Phi^\pm\rangle_{SA} = \frac{1}{\sqrt{2}}(|\uparrow\rangle_S|\uparrow\rangle_A \pm |\downarrow\rangle_S|\downarrow\rangle_A) \tag{5}$$

Express the state $|\psi\rangle_{SAB}$ in terms of the Bell bases $|\Psi^\pm\rangle_{SA}$ and $|\Phi^\pm\rangle_{SA}$, and we obtain:

$$\begin{aligned}|\psi\rangle_{SAB} = &\frac{1}{2}[(-\alpha|\uparrow\rangle_B - \beta|\downarrow\rangle_B)|\Psi^-\rangle_{SA} + (-\alpha|\uparrow\rangle_B + \beta|\downarrow\rangle_B)|\Psi^+\rangle_{SA}] \\ &+ \frac{1}{\sqrt{2}}[(\alpha|\downarrow\rangle_B + \beta|\uparrow\rangle_B)|\Phi^-\rangle_{SA} + (\alpha|\downarrow\rangle_B - \beta|\uparrow\rangle_B)|\Phi^+\rangle_{SA}]\end{aligned} \tag{6}$$

After Alice informs Bob the result of the measurement through classical communication, Bob performs an appropriate unitary transformation on *B* depending on the result. If Alice obtains state

$|\Psi^-\rangle_{SA}$, Bob does nothing; if Alice $|\Psi^+\rangle_{SA}$, Bob performs transformation $\sigma_z = \begin{pmatrix} 1 & 0 \\ 0 & -1 \end{pmatrix}$ on B; if Alice $|\Phi^-\rangle_{SA}$, Bob does $\sigma_x = \begin{pmatrix} 0 & 1 \\ 1 & 0 \end{pmatrix}$; and if Alice $|\Phi^+\rangle_{SA}$, Bob does $\sigma_y = \begin{pmatrix} 0 & -i \\ i & 0 \end{pmatrix}$. According to quantum teleportation protocol, Bob obtains the state:

$$|\psi\rangle_B = \alpha |\uparrow\rangle_B + \beta |\downarrow\rangle_B \tag{7}$$

Thus the coherent superposition state $|\psi\rangle_S$ near to the output is fed back to distant Bob. Then Bob sends the state $|\psi\rangle_B$ into the feedback processor. Using entanglement [14], we add a second quantum spin into the feedback loop as a control spin, initially in the state $|\downarrow\rangle_C$ and make a CNOT operation, and now the two quantum spins are in the state $\alpha|\uparrow\rangle_B|\uparrow\rangle_C + \beta|\downarrow\rangle_B|\downarrow\rangle_C$. Obviously, the state $\alpha|\uparrow\rangle_B|\uparrow\rangle_C + \beta|\downarrow\rangle_B|\downarrow\rangle_C$ exhibits an entanglement correlation, so the CNOT operation has caused the second spin to obtain quantum information about the spin to be controlled. Then we apply a pulse to the system in state $\alpha|\uparrow\rangle_B|\uparrow\rangle_C + \beta|\downarrow\rangle_B|\downarrow\rangle_C$ to flip the first spin if and only if the second spin is up.

Since quantum teleportation can transmit quantum information from Alice's particle to Bob's particle over arbitrary distance [29], by contrast to Lloyd's coherent feedback [14], this scheme also can accomplish some distant quantum feedback control tasks. Although Alice operates Bell-state measurement and destroys the unknown initial coherent state, Bob receives its exact replica, so we may think that the coherence of feedback loop is not destroyed. Evidently, teleportation-based strategy can perform some tasks such as distant feedback of coherent quantum state that can't be completed through such coherent control with classical feedback as Markovian and Bayesian quantum feedback. As noted above, coherent control with classical feedback involves quantum measurement and it irrevocably destroys an unknown initial state of the system, so it is typically stochastic and destructive. In its feedback loop, the quantum coherence can't be protected. However, this quantum feedback control scheme can be thought deterministic, nondestructive and preserving coherence in feedback loop.

## 4. Quantum feedback control with quantum cloning

As an essential idea in classical control, feedback is used to compensate the effects of unpredictable disturbances on a system under control, or to make automatic control possible when the initial state of the system is unknown. To control a system, one must obtain the information about the evolving system state through measurement. However, in quantum feedback control, it is impossible to extract information about the state of system through measurement without disturbing it. Moreover, measurement backaction greatly complicates the notion of quantum feedback control. Following Ting's idea [27], we renounce precise feedback and add an optimal cloning machine at the output side. Different from Ting's complete abandonment of measurement, we separate quantum feedback control design into a state recognition step involving measurement and a feedback control step without measurement. So this scheme establishes a compromise between information acquisition and measurement disturbance. This is obviously different from traditional quantum feedback since there the information acquisition necessarily results in destroying the state of quantum system.

The general picture of quantum feedback control with quantum cloning is as follows. The quantum system to be controlled is called control object. The actuator generates input signal to drive the object and its output is sent into an optimal cloning machine (cloner). The cloner inaccurately clones the state, generates (*N+M+1*) copies, and a copy is taken as the output of system. State recognizer receives *N* unknown copies, makes some measurements on them and obtains some information about their states. Then, one can recognize the *N* copies through appropriate recognition algorithm. If the *N* copies are "good" enough, the recognizer gives an "on" signal to the actuator, and the actuator receives another *M* copies from the cloner as feedback and generates new quantum signal to drive the object until the given control target is reached. Here, the new signal of the actuator is determined by the feedback information and control target. If the N copies are not "good" enough, the recognizer will send an "off" signal to the actuator, and the actuator will not receive feedback information determined by the cloner.

According to no-cloning theorem [33], arbitrary unknown quantum state cannot be copied exactly, and this is also an important difference between quantum control system and classical control system. However, this doesn't exclude the possibility of approximately cloning quantum state [34]. In fact, there is optimal universal quantum cloning machine that can approximately copy arbitrary unknown quantum state with unity probability [35]. Besides approximate cloning, linearly independent quantum states can also be probabilistically cloned by a general unitary reduction operation with probability less than unity [36]. In probabilistic cloning, the machine yields faithful copies of the input state with a postselection of the measurement result. However, optimal universal quantum cloning machine can be decomposed into rotations and controlled NOT gates and doesn't involve measurement. Here, we use approximately cloning to amplitude the output state of the object, that is to say, the output state is approximately copied, and some copies are used to obtain information through quantum measurement and other copies are used to determine feedback control.

The ideal 1→2 cloning process is described by the transformation:

$$|s\rangle_a |M\rangle_x \rightarrow |s\rangle_a |s\rangle_b |\tilde{M}\rangle_x \qquad (8)$$

where $|s\rangle_a$ is the state of the original mode, $|s\rangle_b$ is the copied state, $|M\rangle_x$ is the original state of the quantum cloning machine and $|\tilde{M}\rangle_x$ is the final state of the quantum cloning machine. The whole process of quantum cloning is to produce at the output of the cloning machine two identical states $|s\rangle_a$ and $|s\rangle_b$ in the modes a and b, respectively. Considering 1→*M+N+1* cloning process, it can be described by the transformation:

$$|s\rangle_a |M\rangle_x \rightarrow |s\rangle_a |s\rangle_b \cdots |s\rangle_{M+N+1} |\tilde{M}\rangle_x \qquad (9)$$

The cloner generates (*M+N+1*) copies and sends *N* copies into the state recognizer. Assume that the recognizer receives *N* unknown copies $U_R = \{|P_1\rangle, |P_2\rangle, \cdots, |P_N\rangle\}$, makes some measurements on them and obtains some information about their states.

In quantum mechanics, states of quantum systems are represented by quantum states and quantum states can be represented by the vector $|\phi\rangle$ in Hilbert space:

$$|\phi\rangle = \sum_k a_k |\varphi_k\rangle \qquad (10)$$

where $a_k$ is complex number and satisfies $\sum_k |a_k|^2 = 1$. $\{|\varphi_1\rangle, |\varphi_2\rangle, \cdots, |\varphi_k\rangle\}$ is a set of bases of the vector $|\phi\rangle$ and is called *k* eigenstates of quantum states $|\phi\rangle$, namely all possible states when $|\phi\rangle$ is measured. $|a_k|^2$ represents the occurrence probability of eigenstate $|\varphi_k\rangle$ and the information about

the phase is contained in the argument of complex number $a_k$.

After some measurement on the $N$ unknown copies of the recognizer, one can get some information of unknown copies. Assume that $N(N \geq 1)$ quantum states $|P_s\rangle (1 \leq s \leq N)$ can be respectively represented as follows:

$$|P_s\rangle = \sum_{j=1}^{n_s} \beta_j^{(s)} | p_j^{(s)} \rangle \tag{11}$$

where $s \in \{1, 2, \cdots, N\}$, $\beta_j^{(s)}$ are complex number coefficients and satisfy $\sum_{j=1}^{n_s} |\beta_j^{(s)}|^2 = 1$. $U_s = \{|p_1^{(s)}\rangle, |p_2^{(s)}\rangle, \cdots, |p_{n_s}^{(s)}\rangle\}$ is composed of $n_s$ orthogonal eigenstates of $|P_s\rangle$ and these eigenstates can constitute a set of orthogonal bases in Hilbert space. We use $|U|$ to express the number of elements in the set $U$, then $|U_s| = n_s$. Let $\tilde{U}$ represent the set of the extended bases [37]:

$$\tilde{U} = U_o \cup U_1 \cup U_2 \cup \cdots \cup U_N = \{|\tilde{p}_1\rangle, |\tilde{p}_2\rangle, \cdots, |\tilde{p}_l\rangle\} \tag{12}$$

that is to say, define the union of these eigenstate sets as the set of extended bases. The elements in the set of extended bases are not necessarily orthogonal. Now express the all quantum states with the extended bases $\{|\tilde{p}_1\rangle, |\tilde{p}_2\rangle, \cdots, |\tilde{p}_l\rangle\}$:

$$|\tilde{P}_s\rangle = \sum_{k=1}^{l} \tilde{\beta}_k^{(s)} |\tilde{p}_k\rangle \tag{13}$$

where if $|\tilde{p}_k\rangle$ is equal to $|p_j^{(s)}\rangle$, $\tilde{\beta}_k^{(s)}$ is equal to $\beta_j^{(s)}$; otherwise $\tilde{\beta}_k^{(s)}$ is zero. Then calculate their arithmetical mean value $|\overline{P}\rangle$:

$$|\overline{P}\rangle = \sum_{k=1}^{l} (\frac{1}{T} \sum_{s=1}^{T} \beta_k^{(s)}) |\tilde{p}_k\rangle = \sum_{k=1}^{l} \tilde{\alpha}_k |\tilde{p}_k\rangle \tag{14}$$

Regard $|\overline{P}\rangle$ as the "objective" state. Based on the above notations, we may define state-distance between the given quantum states $|P_s\rangle$ and the "objective" quantum state $|\overline{P}\rangle$:

**Definition 1:** State-distance between $N(N \geq 1)$ given quantum states $|P_s\rangle (1 \leq s \leq N)$ expressed by Eq.(11) and the "objective" state $|\overline{P}\rangle$ expressed by Eq.(14) is defined as:

$$d(|P_s\rangle, |\overline{P}\rangle) = (\sum_{k=1}^{l} |\tilde{\beta}_k^{(s)} - \tilde{\alpha}_k|^2)^{\frac{1}{2}} \tag{15}$$

where $\tilde{\beta}_k^{(s)}$ and $\tilde{\alpha}_k$ are corresponding complex numbers in Eq.(13) and Eq.(14), respectively.

According to the definition of state-distance, it is obvious that $d(|P_s\rangle, |\overline{P}\rangle) \geq 0$ and $d(|P_s\rangle, |\overline{P}\rangle) = 0$ if only if $|\overline{P}\rangle$ and $|P_s\rangle$ are the same, that is to say, state-distance will reach the minimum zero when all given quantum states are identical.

Calculate $d(|P_s\rangle,|\overline{P}\rangle)$ according to Eq.(15) respectively and compare them with the beforehand given state-distance threshold $d_0$. Here, the state-distance threshold should be selected according to the need of specific problem. If all state-distances satisfy $d(|P_s\rangle,|\overline{P}\rangle) < d_0$, recognize them as a class, the state recognizer generates an "on" signal, and the actuator receives another *M* copies from the cloner as feedback. Considering the feedback information and comparing the output with the control target, the actuator generates new quantum signal to drive the object. Otherwise, the state recognizer will send an "off" signal to the actuator, and the actuator will not receive feedback copies from the cloner.

In this feedback strategy, the problem of designing quantum feedback algorithms is separated into two steps: a state recognition step and a feedback control step. The aim of state recognition step is to obtain information and the process involves quantum measurement, so it is destructive. However, the feedback process doesn't necessarily acquire information and doesn't involve measurement, so it can preserve quantum coherence. In fact, we give a compromise between information acquisition and measurement disturbance in view of the characteristics of measurement in quantum control. The compromise is realized through approximately cloning and renouncing precise feedback. Since the coherence of feedback information can be preserved, this feedback strategy can perform coherent feedback control at the cost of feeding precisely back the output.

## 5. Conclusion

Quantum control theory and quantum information technology are two rapidly developing new subjects. With the gradual maturation of quantum information technology, it is possible that quantum control theory might profitably be reexamined from the perspective of quantum information. Following this idea, this paper presents two results on the role of quantum information technology in quantum feedback control and proposes two quantum feedback control schemes: one is teleportation-based distant quantum feedback control and the other is quantum feedback control with quantum cloning. In the first scheme, the output from the quantum system to be controlled is fed back into the actuator via teleportation with preserving the coherence of feedback information to alter the dynamics of system. We theoretically shows that feedback controller including a teleportation procedure can accomplish some tasks such as distant feedback quantum control that Markovian or Bayesian quantum feedback can't complete. In the second quantum feedback strategy, the design of quantum feedback control algorithms is separated into a state recognition strategy, which gives "on-off" signal to the actuator through recognizing some copies from the cloning machine, and a feedback (control) strategy with another copies of cloning machine. The recognition process involves quantum measurement and is destructive; however, the feedback process is preserving quantum coherence, so this scheme can perform some quantum control tasks with coherent feedback. Besides quantum control, this two quantum feedback schemes also have important potential application to large-scale quantum computation, quantum communication networks and quantum robots [38].

## Acknowledgements

This work has been supported by the Innovation Research Project of Science and Society Practice of the Chinese Academy of Science for Graduate Students (No. 2004-18), and the Innovation Fund of the University of Science and Technology of China for Graduate Students (KD2004048).